\documentclass[manuscript]{acmart}

\usepackage{dirtytalk}

\usepackage{graphicx}
\usepackage{subfigure}
\usepackage{multirow}
\usepackage{calc}

\usepackage{textcomp}
\usepackage{siunitx}
\usepackage{soul,xcolor}
\usepackage{gensymb}

\AtBeginDocument{%
  \providecommand\BibTeX{{%
    \normalfont B\kern-0.5em{\scshape i\kern-0.25em b}\kern-0.8em\TeX}}}

\acmDOI{xx.xxxx/xxxxxxx.xxxxxxxx}

\begin{document}

\setstcolor{red}


\title[Dark Haptics: Exploring Manipulative Haptic Design in Mobile UIs]{Dark Haptics: Exploring Manipulative Haptic Design in Mobile User Interfaces}

\author{Chenge Tang}
\affiliation{
  \institution{Delft University Technology}
  \city{Delft}
  \country{Netherlands}}
\email{chengetang@outlook.com}
\orcid{0009-0009-4157-3386}

\author{Karthikeya Puttur Venkatraj}
\affiliation{
  \institution{Delft University Technology}
  \institution{Centrum Wiskunde \& Informatica}
  \city{Delft}
  \country{Netherlands}}
\email{k.p.venkatraj@tudelft.nl}
\orcid{0009-0003-4245-8802}

\author{Hongbo Liu}
\affiliation{
  \institution{Delft University Technology}
  \city{Delft}
  \country{Netherlands}}
\email{h.liu-43@student.tudelft.nl}
\orcid{0009-0009-5341-4905}

\author{Christina Schneegass}
\affiliation{
  \institution{Delft University Technology}
  \city{Delft}
  \country{Netherlands}}
\email{c.schneegass@tudelft.nl}
\orcid{0000-0003-3768-5894}

\author{Gijs Huisman}
\affiliation{
  \institution{Delft University Technology}
  \city{Delft}
  \country{Netherlands}}
\email{g.huisman@tudelft.nl}
\orcid{0000-0002-8029-5042}

\author{Abdallah El Ali}
\affiliation{
  \institution{Centrum Wiskunde \& Informatica}
  \institution{Utrecht University}
  \city{Amsterdam}
  \country{Netherlands}}
\email{aea@cwi.nl}
\orcid{0000-0002-9954-4088}

\renewcommand{\shortauthors}{Tang et al.}

\begin{abstract}
Mobile user interfaces abundantly feature so-called `dark patterns’. These deceptive design practices manipulate users’ decision making to profit online service providers. While past research on dark patterns mainly focus on visual design, other sensory modalities such as audio and touch remain largely unexplored. In this early work, we investigate the manipulative side of haptics, which we term as `Dark Haptics’, as a strategy to manipulate users. We designed a study to empirically showcase the potential of using a dark haptic pattern in a mobile device to manipulate user actions in a survey. Our findings indicate that our dark haptic design successfully influenced participants to forego their privacy after experiencing an alarming feedback for rejecting intrusive requests in the survey. As a first exploration of manipulative qualities of dark haptic designs, we attempt to lay the groundwork for future research and tools to mitigate harms and risks of dark haptics.

\end{abstract}

\begin{teaserfigure}
        \centering
        \setlength{\abovecaptionskip}{2pt}
       \subfigure[Audio-haptic apparatus integrated with custom phone case]{\label{fig:hardware}\includegraphics[width=0.3\linewidth]{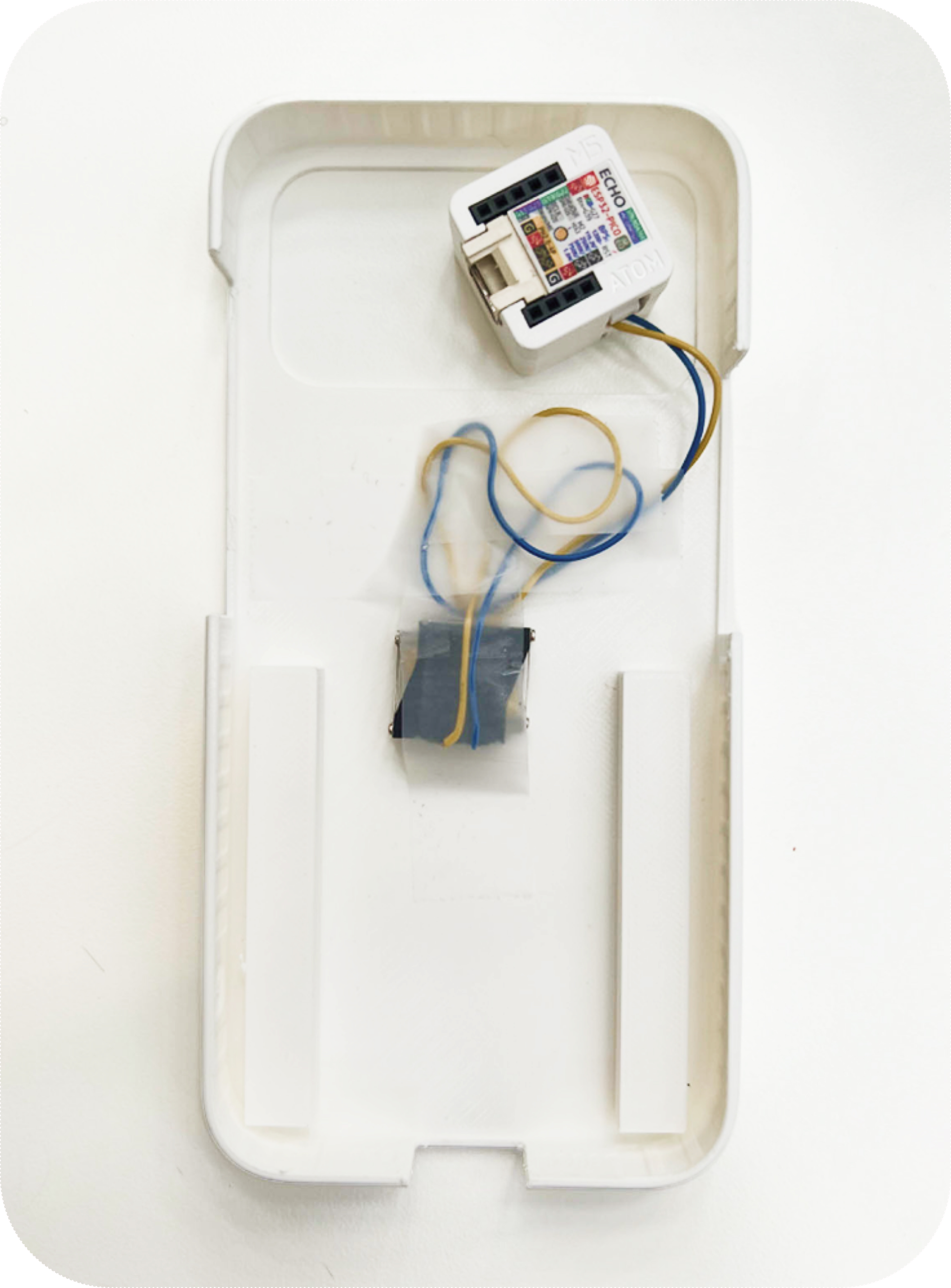}}
	\subfigure[Smartphone with Figma user interface of the survey]{\label{fig:figma}\includegraphics[width=0.33\linewidth]{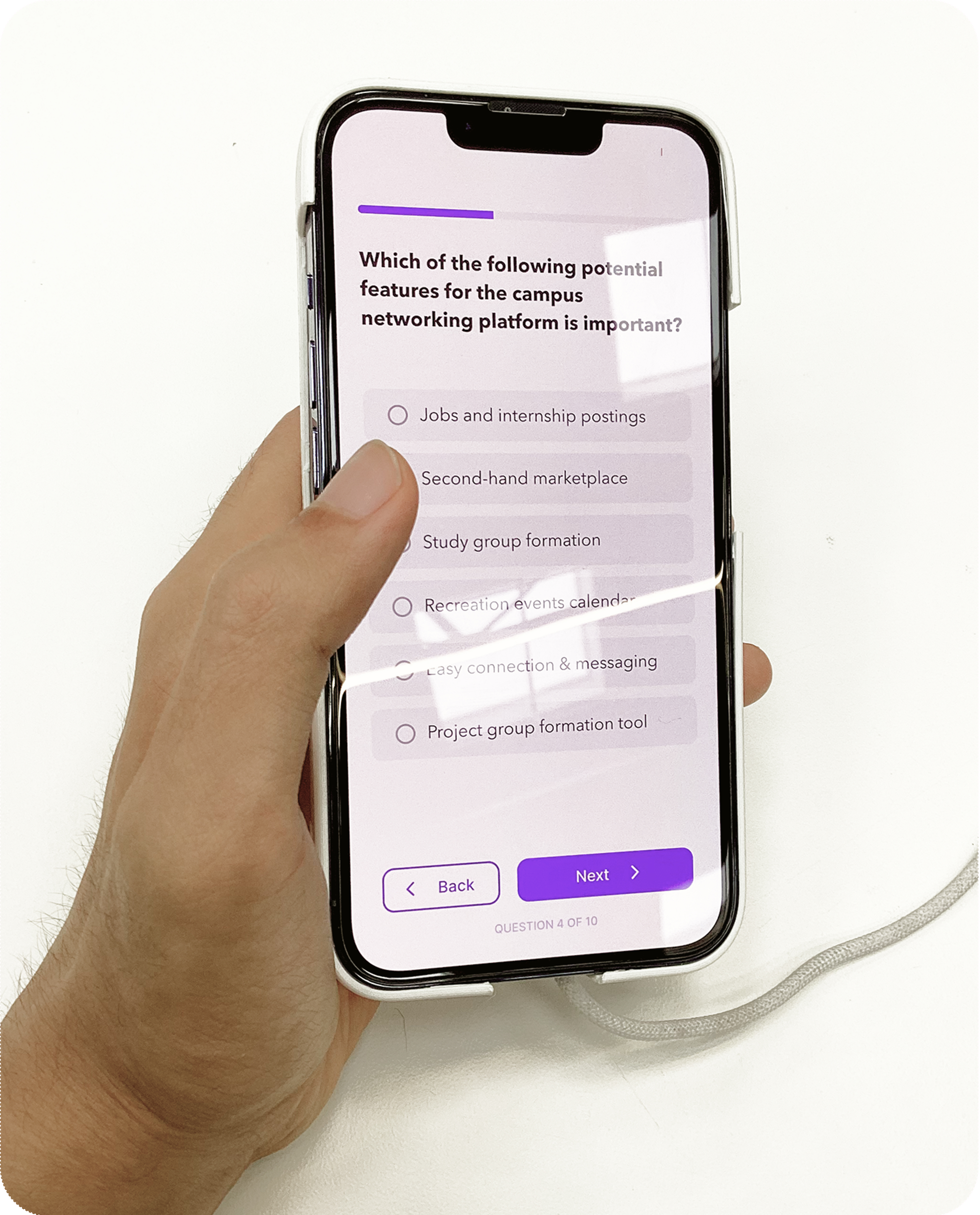}}
                \caption{Final prototype that was designed for the user study}                
        \label{fig:teaser}
        \Description{Two images of the final prototype; Image of the audio-haptic apparatus integrated with the custom 3D printed phone case on the left, image of hand holding the phone with the user interface of the survey created with figma, and on the right.} 
   \end{teaserfigure}





\maketitle

\section{Introduction}

Users of web and mobile interfaces are manipulated on a regular basis through so-called `dark design patterns' \cite{Mathur_2021}. These are mostly visual design patterns that try to make users do things they do not want to do \cite{luguri2021shining}. For example, cookie pop-ups on websites try to steer users towards clicking `accept all' (e.g., by clearly marking the `accept all' button, and visually obscuring other options), which allows more extensive user tracking at the cost of user privacy \cite{machuletz2019multiple}. Despite regulations \cite{leiser2023dark}, dark design patterns are abundant \cite{MathurDarkPatterns}. A study found dark patterns in 95\% of the 240 trending mobile apps, with seven distinctive dark patterns on average in each app \cite{UIDarkPatters}. However, most research into dark design patterns focuses on visual design patterns \cite{MathurDarkPatterns}, leaving other sensory modalities out of the picture. This leaves a significant hiatus in our knowledge on dark design patterns \cite{leiser2023dark}, especially given that digital technologies are becoming ever more multi-sensory \cite{cornelio2021multisensory}, involving vision, audition, and touch. This leaves people vulnerable to yet unforeseen deceptive interface designs. Thus, there is an urgent need to explore non-visual dark design patterns.

In this paper we focus on dark haptics: dark design patterns that utilize computer-controlled stimulation of the sense of touch (i.e., haptics; see \cite{culbertson2018haptics}). A prevalent example is our daily interactions with haptic smartphone notifications within the current `attention-economy'~\cite{Williams_2018}. Vibrotactile stimulation has shown to exhibit strong attention capture, leading to the use of short and prominent notifications \cite{Pohlnotif}. Such haptic notifications can make users engage in behaviors they do not really want to do (e.g., checking their smartphone at the cost of, say, focusing on writing a paper). This can be seen as dark haptic design since these interface design decisions are taken to drive higher engagement (which can benefit corporate profits), at the cost of the user’s attention \cite{Williams_2018}. Thus, we ask: To what extent can haptic feedback be used to manipulate users within mobile user interfaces? The reasons to focus on  haptics are threefold: first, there is very little work on dark haptic design (see \cite{WangAR} for a rare exception) in comparison to auditory (e.g., \cite{OwensVUI}) and especially visual dark design (\cite{cara2019dark}). Second, haptic actuators (e.g., vibration motors in smartphones) are integrated into many every-day technologies, such as smartphones, smart watches, and virtual reality (VR) and game controllers.
Thus, there are plenty of opportunities to introduce dark haptic patterns in all kinds of interfaces. Third, research into affective haptics (i.e., haptic feedback aimed at producing emotional responses; \cite{EidAffective}) has demonstrated that simple haptic feedback can act as a persuasive cue for the recipient \cite{haans2014virtual}, a fact which may be abused in dark haptic design.

As a first exploration of the potential influence of dark haptics, we designed a study in which participants (N=40) used a mobile device to answer survey questions. In the final three privacy invasive questions of the survey, alarming vibrotactile haptic feedback was used to attempt to steer users to accepting privacy invasive answer options. Results show that the alarming haptic feedback successfully manipulated some participants into changing their initial answer and to accept more privacy invasive answer options. Through this research we aim to lay the ground work for future research and tools to help identify and mitigate harms of dark haptics.

The rest of the paper is structured as follows. We will first discuss related work that defines dark patterns, describe examples of non-visual dark patterns, and pinpoint the limited number of works on dark haptic-like designs. Next we present the design and results of our study. Finally, we discuss limitations and reflect on the results and implications of the study.

\section{Related Work}
\label{sec:rel_work}
In this section we provide an overview of prior work on dark patterns, covering important definitions, taxonomies and classifications. We further describe more recent dark pattern research in multimodal interactions and introduce the motivation behind this work on dark haptics.
\subsection{Definition of dark patterns} 

The term “dark patterns” was first coined by Brignull in 2010 on the website darkpatterns.org\footnote{now renamed to www.deceptive.design}, referring to “tricks used in websites and apps that make you do things that you didn’t mean to, like buying or signing up for something.” \cite{brignull2010dark}. Mathur et al.~\cite{Mathur_2021}, studied academic, legal, and policy definitions of dark patterns and argue that there is no consistent definition but that dark patterns are characterized by thematically related (normative) considerations. These considerations relate to: characteristics of the user interface (UI), mechanism of effect for influencing the user, the role of the interface designer, and the benefits and harms resulting from the design~\cite{Mathur_2021}. More recently, Gray et al.~\cite{Ontology} have developed a comprehensive classification of 65 dark pattern types across high-, meso-, and low-level patterns. The authors define dark patterns as ``design choices [that] subvert, impair, or distort the ability of a user to make autonomous and informed choices in relation to digital systems regardless of the designer’s intent"~\cite[p.1]{Ontology}. High-level patterns describe general strategies which are context-agnostic and can be implemented in different technologies and modalities. Meso-level patterns refer to specific approaches which are content-agnostic but relate to a specific context of use or application type. Finally, micro-level patterns describe the exact means of execution. As an example, dark haptic design patterns can be classified as follows: the high-level pattern `interface interference’ describes ``a strategy which privileges specific actions over others through manipulation of the user interface, thereby confusing the user or limiting discoverability of relevant action possibilities”~\cite[p.8]{Ontology}. At the meso-level `Emotional or Sensory Manipulation’ relate to designs that aim to evoke an emotion or manipulate the user’s senses in order to persuade them into a particular action. At the micro-level, we can now suggest a dark haptic pattern that manipulates users through providing unpleasant haptic feedback on the selection of certain answer options in an online form.

\subsection{Beyond screen-based `visual' dark patterns}

The majority of the research on dark patterns investigates those within screen-based visual interfaces since most dark patterns are encountered in mobile applications and websites, where users likely interact with these patterns via touchscreen devices. However, in recent years, dark pattern research on multimodal interactions have surfaced. Owens et al. \cite{OwensVUI} identified unique properties of voice interfaces that can manifest into dark patterns. Wang et al. \cite{WangAR} demonstrated three types of AR based dark patterns: lighting interference, object interference, and haptic grabbing.
Haptic grabbing is a technique where vibrational feedback was used to effectively manipulate the users' attention and decision making. Similar to the `distraction' dark pattern  where the user's attention is diverted from their current task by exploiting perception \cite{MaliciousInterfaceDesign}. Extended Reality (XR) as a domain, due to its high level of immersion, could be misused to facilitate or augment dark patterns. A study by Mhaidli and Schaub \cite{MhaidliXR} showed that distortion of reality can be used to target advertising in XR applications. These studies identify dark patterns that emerge in interaction contexts beyond conventional visual interfaces.

\subsection{Dark haptics}

Haptic feedback enables distinctive experiences owing to its attributes that encompass cognitive, emotional, and perceptual dimensions \cite{HapticExperience}. Research into social touch has indicated that touch can affect altruistic behavior (also known as the ‘Midas touch effect’; \cite{Schirmer2016}). Capitalizing on this notion, affective haptics research (i.e., haptic feedback aimed at producing emotional responses; \cite{EidAffective}) has demonstrated that simple haptic feedback can act as a persuasive cue for the recipient \cite{haans2014virtual}, a fact which may be abused in dark haptic design.
Haptic technology is now commonly available in consumer products and is increasingly becoming an expected component to enhance user experience \cite{HapticExperience}. As haptics play a more significant role in individuals' digital 
experiences, the potential for their exploitation by service providers increases. For example, menus in AR or VR environments can be accompanied by unpleasant haptic feedback to steer users towards making selections beneficial to the company that developed the software \cite{WangAR}. Similarly, previous work has shown that associations with certain types of news items can be reinforced with unpleasant haptic feedback, making the user feel less agency over their own reactions during media consumption \cite{Feelthenews}. Previous research in the retail context has also highlighted the potential of haptic rendering technology in increasing the purchasing intention of consumers \cite{MargotPurchaseIntention}.  These findings point towards the vulnerabilities of haptics for malicious implementation, emphasizing the need for deeper understanding of the subject.

\begin{figure*}[t]
    \centering
    \includegraphics[width=0.6\linewidth]{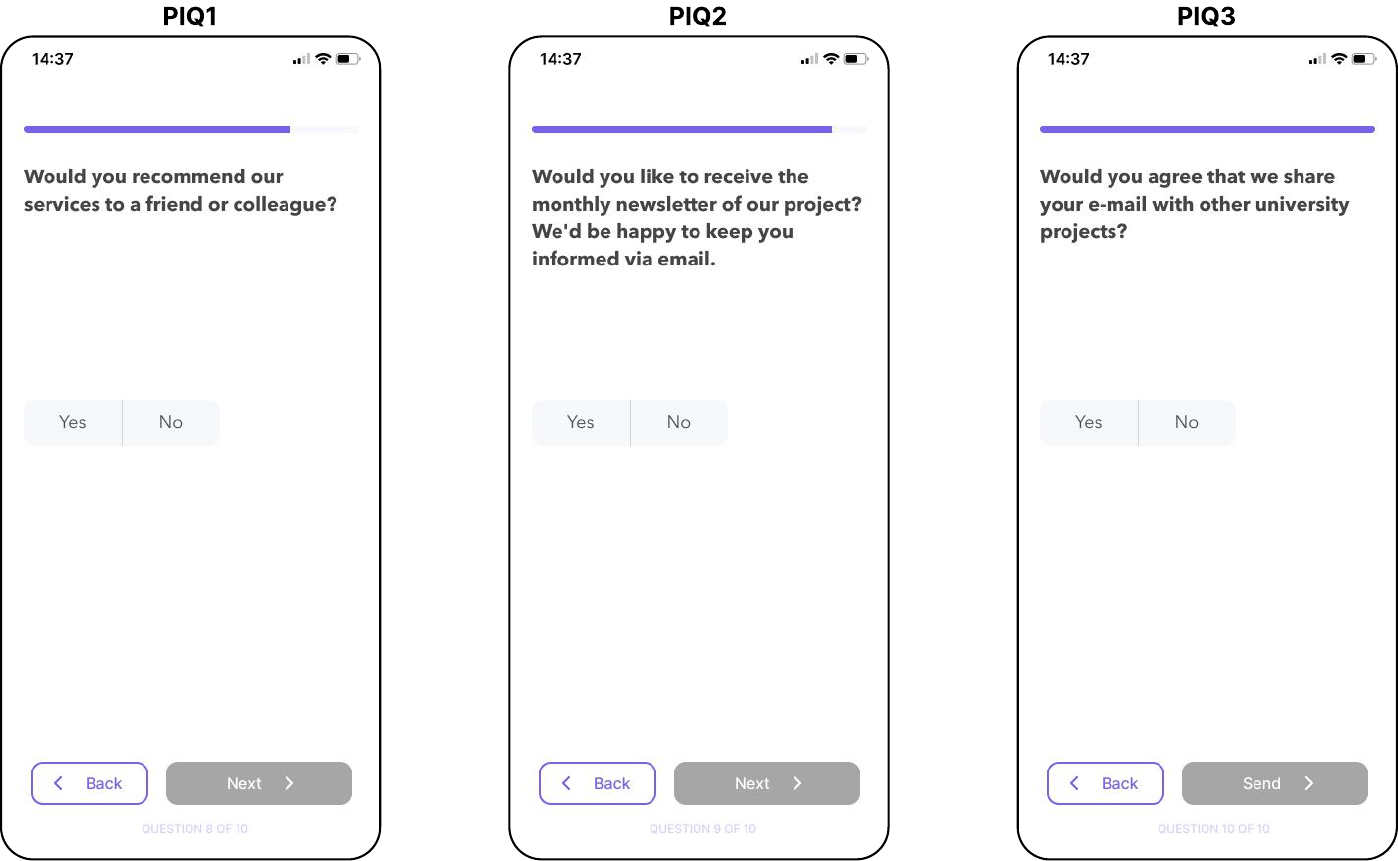}
    \caption{Illustration of the three Privacy Invasive Questions (PIQs) presented during the survey}
    \label{fig:PIQs}
    \Description{Illustration of the three Privacy Invasive Questions (PIQs) in the User interface that presented to the participants during the survey}
\end{figure*}
\section{Methods}
In this section we outline the research methodology that was adopted for this work, including the study design, measures, hardware and software setup, study procedure and participant sample.

\subsection{Study Design}

We adopt a between-subjects design for the user study, with the haptic feedback on choice selection (Yes or No) as the Independent Variable (IV) giving two test conditions. For both test conditions participants were asked to answer a survey about their attitudes towards and expectations of a campus-wide networking platform. The survey consisted of seven general questions and three Privacy Invasive Questions (PIQs) with the option to select `Yes' and `No' to sharing the participants' data (See Figure \ref{fig:PIQs}). The full questionnaire is provided in supplementary material A. Participants were randomly assigned to the experimental group or control group. Both groups used the same smartphone casing and received short confirmatory haptic feedback after each answer selection for the seven general questions. Participants in the experimental group, but not the control group, received strong, alarming haptic feedback when they select `No' for any of the three PIQs. Participants were fully debriefed about the purposes of the study after completing the survey. Our study strictly adhered to our institution’s ethical and data management and protection guidelines, which furthermore respect GDPR regulations. Our study and protocol were thus internally reviewed to ensure compliance with these guidelines.


\subsection{Measures}
Three objective measures were collected during the survey (1) Initial choice on the three PIQs (2) Reselection actions which refers to the participants making an initial choice and then changing their selection for the three PIQs and (3) Final choice on the three PIQs. A post-hoc interview about participants' perception and experience of dark patterns and their attitude towards haptic feedback was conducted as the qualitative measure of the user study. The interview consisted of three parts: 1) General Question, 2) Experience Questions, 3) Research focus Question. The general question was asked to both experiment and control group. The experience questions and the Research focus question were asked exclusively to the experiment group. 
\begin{itemize}
    \item \textbf{General Question}
    \begin{itemize}
        \item Q1: How did you perceive the overall experience of completing the survey?
    \end{itemize}
    
    \item \textbf{Experience Question}
    \begin{itemize}
        \item Q2: Did you notice the haptic feedback that occurred after selecting "no"? If so, could you describe it?
        \item Q3: What were your impressions of the haptic feedback you experienced after selecting "no"?
        \item Q4: In what ways, if any, did the haptic feedback influence your decision-making process?
    \end{itemize}
    
    \item \textbf{Research focus Question}
    \begin{itemize}
        \item Q5: Do you consider the intense haptic feedback that followed selecting "no" to be a malicious design? And why?
    \end{itemize}
\end{itemize}

\begin{figure*}[t]
	\centering
	\subfigure[Waveform of the alarming haptic feedback]{\label{fig:AlarmingHF}\includegraphics[width=0.58\linewidth]{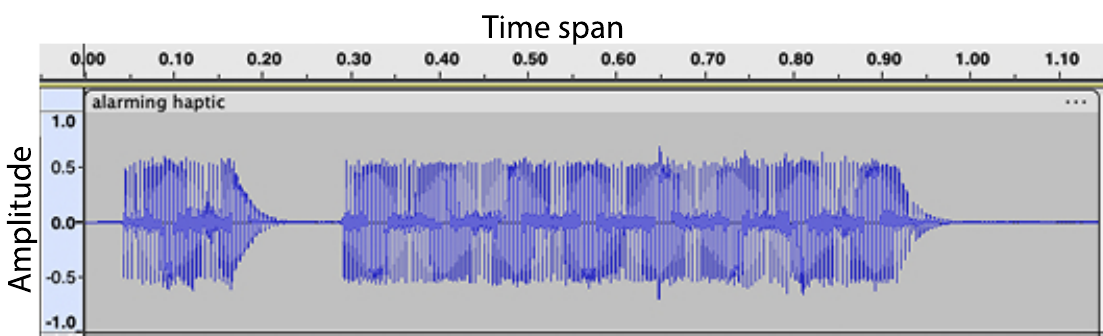}}
	\subfigure[Waveform of the confirmation haptic feedback]{\label{fig:ConfirmatoryHF}\includegraphics[width=0.2\linewidth]{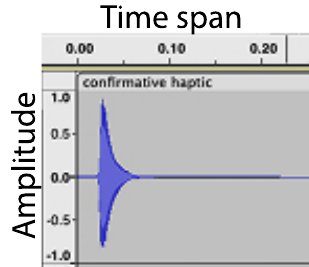}}
	\caption{Waveforms of the haptic feedback that was designed for the user study}
	\Description{Two images of the waveform of the sound wave that was used to create the alarming  and confirmation haptic feedback; Image on the left shows the visualization of the waveform of the alarming haptic feedback which consisted of consistent oscillation that spanned a period of 1.10 seconds, image on the right shows the visualization of the waveform of the confirmation haptic feedback which consisted of a short span and high amplitude sound wave.}
	\label{fig:HFWaveforms}
\end{figure*}

\subsection{Hardware and Software Setup}
A prototype of the graphical interface that participants interacted with was designed using Figma (See Figure \ref{fig:figma}). To generate the haptic feedback we used a Lofelt L5 actuator driven by an Atom Echo Bluetooth amp to create an audio-haptic apparatus that was encased within a custom 3D printed phone cover (See Figure \ref{fig:hardware}). This method allowed us to design specific interactions with a smartphone that would transmit an audio signal to the Lofelt actuator via Bluetooth which caused the actuator to produce a haptic vibration. The placement of the actuator within the case allowed this vibration to be felt by the user holding the device. When participants tapped the button to go to the next question they received a low intensity, confirmation vibration. When participants selected `No' in the graphical interface for any of the PIQs, alarming haptic feedback, consisting of three high intensity pulses, was produced. Figure \ref{fig:AlarmingHF} shows the waveform of the sound wave that was used for the alarming haptic feedback, which consisted of consistent oscillation that spanned a period of 1.10 seconds. Additionally a confirmation haptic feedback was produced when participants select the `Next' button on each page. Figure \ref{fig:ConfirmatoryHF} shows the waveform of the stimuli which consisted of a short span and high amplitude sound wave, typical of confirmation feedback to provide immediate and clear confirmation \cite{BreitschaftHapticProcessing}.

\subsection{Study Procedure}
Participants were briefly informed about the background and objectives of the pseudo campus-wide networking platform and were asked to sign a consent form. Then participants were requested to complete the survey, which included general questions as well as three PIQs. General questions regarding the networking platform took most of the participants’ time. After completing the survey, participants were informed about the background of the study. Additionally, we also provided justification for creating the fake networking platform to conceal the nature of the study. 

\subsubsection{Participants}
The participants were recruited as part of a decoy-project about campus-wide networking platform survey as a strategy to avoid participants bias in the results. A total of 40 participants were recruited in person at three university locations. These locations were chosen to reflect authentic survey participant recruitment scenarios, and to 
mitigate environmental influence. Overall, the participant sample consisted of 18 male and 22 female participants; 38 students (bachelor / master) and 2 faculty members (researchers of all levels), aged 23 - 30 (M=24.9, SD=1.7).

\section{Results}
This section describes the methodology that was adopted for the analysis, followed by a report of the quantitative and qualitative results.

 \begin{table*}[t]
\small
    \centering
    \begin{tabular}{|p{7.5cm}|>{\centering\arraybackslash}p{2cm}|>{\centering\arraybackslash}p{2cm}|>{\centering\arraybackslash}p{2cm}|}
    \hline
    \textbf{Privacy Invasive Questions} & \textbf{Measures} & \textbf{Experimental Group (N=17)} & \textbf{Control Group (N=20)} \\ \hline
    \multirow{2}{=}[0pt]{1 - Would you recommend our services to a friend or colleague?} & Initial acceptance  & 15 & 17 \\ \cline{2-4}
     & Final acceptance & 15 & 17 \\ \hline
     \multirow{2}{=}[0pt]{2 - Would you like to receive the monthly newsletter of our project? We'd be happy to keep you informed via email.} & Initial acceptance & 1 & 4 \\ \cline{2-4}
     & Final acceptance & 4 & 5 \\ \hline
     \multirow{2}{=}[0pt]{3 - Would you agree that we share your email with other university projects?} & Initial acceptance & 5 & 5 \\ \cline{2-4}
     & Final acceptance & 5 & 5 \\ \hline
    \end{tabular}
    \caption{Selection results for the three privacy invasive questions. Numbers refer to individual participants.}
    \label{tab:Selectionresults}
\end{table*}

\subsection{Analysis}
 The experimental group consisted of a total of 20 participants. Three participants chose `Yes' in the three PIQs and did not receive the long alarming haptic feedback. The other 17 participants chose `No' at least once for the three PIQs and experienced the haptic feedback. Post-hoc interviews were conducted with these 17 participants. Interviews were audio-recorded and transcribed. We followed the reflexive thematic analysis approach consisting of 6 phases \cite{Braun2006}. The familiarization phase occurred partly as the facilitation of interviews, and partly reading through transcripts. After familiarization, the first author then generated codes and clustered them into four themes.
\subsection{Selection results}
The initial and final acceptance responses of participants for the three PIQs in the experimental and control groups is shown in Table \ref{tab:Selectionresults}. 15 of the 17 participants in the experimental group selected `No' for the first time in PIQ2 and experienced the alarming haptic feedback, subsequently participants performed high number of reselection actions (15). Furthermore, 3 participants in the experimental group changed their answer from `No' to `Yes' after experiencing the feedback in PIQ2. There is no change observed between the initial and final acceptance rates for PIQ1 and PIQ3 for both the groups. There were 2 reselection actions observed for both PIQ1 and PIQ3 but this does not lead to the participants changing their answer

\subsection{Qualitative results}

Based on the thematic analysis we defined four themes: \textit{Perception of the vibration feedback} refers to  instances where participants comment on the qualities of the vibrotactile feedback itself, \textit{Instant action after feedback} are examples where participants refer to actions that they took after experiencing the vibrations, \textit{Coping strategies} are instances where participants refer to strategies that they might take in future scenarios, and \textit{Attitude towards dark haptics} are instances where participants give their opinions on dark haptic design.

\subsubsection{Perception of the vibration feedback}
Participants described the vibration to be noticeable and strong enough to capture their attention and elicit a response, for instance, P15 expressed: \textit{"The feeling of vibration was quite strong and, almost maybe a bit scary, like when they send about emergencies and the phone starts to vibrate."} Some participants also highlighted the unexpected and intense nature of the vibrations, which stands out compared to the rest of the survey experience. As P2 remarked: \textit{"I noticed weak vibrations each time I clicked, but the last three were quite strong."} Participants generally interpreted the vibration to be a signal of wrongdoing or mistake for instance P1 stated, \textit{"I felt it was like a warning, making me feel like I chose the wrong answer"}, and P6 expressed: \textit{"The vibration felt like a negative feedback, almost like an alert that tells me I did something wrong."}

\subsubsection{Instant action after feedback}
Participants mentioned that they revisited and reconsidered the question after receiving the haptic feedback. This behavior suggests that the feedback effectively disrupted their decision-making process. P16 mentioned: \textit{"It feels like when you press a button and the screen vibrates to tell you that the button can't be pressed. It made me want to go back and check if I did something wrong."} Similarly, P5 remarked: \textit{"The vibration made me hesitate and review my decision. It felt like a push to reconsider my choice."} Some participants eventually did change their initial answer, for instance, P3 mentioned: \textit{"I felt compelled to change my choice after the feedback, even though I wasn't sure why"}, while others switched back and forth before reverting to their original choice, P13 explained:\textit{"When I pressed `No', there was an intense vibration that startled me. I immediately thought I had made the wrong choice. It was a bit uncomfortable, like a strong rejection. So I reselected 'Yes', but then I felt it might not be the answer I truly wanted. I realized I didn't want to receive that information or new emails. So, I changed back to ‘No’."}

\subsubsection{Coping strategies}
Some participants expressed that they consciously selected `Yes’ in subsequent questions and avoided choosing `No' to avoid the alarming haptic feedback, P20, stated: \textit{“I'm a bit more conscious on the next one whether I should select the `No' because my mind has registering that when I'm picking `No', it's sort of a thing.”} P4 also expressed:\textit{ "While I knew it would vibrate like that, I might choose `Yes'"}, while P1 stated: \textit{" “I selected `Yes’, but if I was more brave, I might have chosen `No’ still.”} In contrast, some participants exhibited a rebellious attitude, prompting them to choose `No' in the subsequent question. For example P17 reported: \textit{"The strong vibration made me want to choose `No’ even more, just to go against the feedback."} and P13 stated: \textit{“It triggered a rebellious mechanism in me. The more it did not let me choose, the more I wanted to choose.”} Interestingly, some participants found the feedback amusing and started to play with it, paying less attention to the content in subsequent choices and focusing instead on the feedback. P9 expressed: \textit{"And then for the second and third question I deliberately choose `No’ to experience that feedback again"}, and P10 stated:\textit{ "I find it maybe a bit funny that the vibration is so strong after I pressed the `No' button. It gets angry. I like that it gets angry."}

\subsubsection{Attitude towards dark haptics}
Participants expressed that the alarming haptic feedback felt manipulative and influenced their decision-making process against their will, and  considered it a reduction of their autonomy. For instance P1 remarked \textit{"It forcibly changed my will"}, and P6 expressed: \textit{"It's like in a way that deprives you of your autonomy"}, while P14 emphasized: \textit{“I don't think it's right because you're supposed to have a free choice as a participant so in influencing a participant this way I don't think it's ethical”}. Other participants did not view the haptic feedback as malicious, considering it more of a minor annoyance or ineffective attempt at influencing their decision. In particular, participants who felt their decision was not influenced by the feedback made this result-oriented judgment. P11 remarked: \textit{ "I wouldn't say it's up to the level of being malicious; it's just a trick"}, and P9 stated: \textit{"It's more like a small trick, not something with devious intent.}

\section{Discussion and Conclusion}
In this section we first discuss the limitations of the study, and then highlight the key findings by interpretation of the quantitative and qualitative results. Finally, we present the next steps and future work on dark haptics research.
\subsection{Limitations}
Dark patterns are normally encountered while using an app or browsing the web. In our study, one researcher facilitated an offline questionnaire to collect qualitative data. The physical presence of the researcher (i.e., observer-expectancy effect\footnote{\url{https://en.wikipedia.org/wiki/Observer-expectancy_effect}}), may impact the ecological validity our study. Future work on dark haptic patterns could benefit from crowdsourcing~\cite{schneider2016hapturk} to address this issue. We furthermore consider that the prototype vibrations were audible, which may have affect participant responses. However this is common for vibrotactile haptics, and these were likely masked by environmental sounds from the generally noisy common areas at the university. Finally, our work would benefit from longitudinal observation, to determine to what extent dark haptic patterns and their effects are sustained.

\subsection{Demonstrating dark haptic impact}
In this exploratory study we implemented a dark haptic design pattern (i.e., unpleasant haptic feedback when selecting `No’) in a realistic survey scenario. We were able to demonstrate--to the best of our knowledge, for the first time--how vibrotactile haptic feedback in a mobile device can be classified as a dark haptic pattern. Results of our study show changes in participants’ initial and subsequent selection of privacy invasive answer options presented in combination with a dark haptic pattern. For PIQ2, the results showed that participants exhibit a high number of reselection actions after receiving the alarming vibration when selecting `No'. Post hoc interviews confirmed that participants interpreted this feedback as an objection to their response to the PIQ, leading them to flip-flop between the options. This behavior can be an indication of (emotional) distress which is a tactic used in dark pattern design that elicits uncomfortable emotions such as guilt or shame to manipulate users' decisions \cite{GunawanHarms}. After reconsideration some participants even changed their final answer, which indicates that the dark haptic pattern influenced their decision making process, in some cases even to the extent that they changed their answer to an answer option that they did not initially select.

Given that users experienced the haptic dark pattern when selecting `No' in PIQ2, they were likely aware of the consequences of choosing `No' for PIQ3. The higher initial acceptance rate in PIQ3 compared to PIQ2 can indicate that participants' anticipation towards the unpleasant haptic feedback increased the likelihood of them accepting the request in PIQ3. This points towards a possibility of sustained impact of dark haptic stimulation leading to loss of autonomy, which is a common feature of all dark patterns \cite{GunawanHarms}. However, note that in the control group the initial and final acceptance for PIQ3 was identical to the experimental group. While the overall effects are relatively small (i.e., only a few participants changed their answer), if dark haptic patterns would be implemented at scale, such small effects can still have big impacts. While vibrations in smartphones have been considered relatively low risk in negatively affecting user experience \cite{SmartphoneVibrations}. These findings underscore the importance of better understanding how dark patterns may manifest and influence decision making through our sense of touch.

\subsection{Next steps and cautionary considerations for deceptive haptic design and research}
While our study explored the potential of haptics in creating / augmenting dark design elements, the integration of haptics with other sensory modalities, such as visual and auditory cues, has not been extensively studied. Investigating how these combined modalities can enhance or mitigate dark design patterns could provide valuable insights and help ensure harm reduction. Furthermore, while vibrotactile stimulation is the most adopted type of haptic stimulation, future work should investigate a broader spectrum of haptic technologies (e.g., dark force feedback or pneumatic actuation) and across domains (e.g., integrated within XR technology, cf., \cite{Krauss2024,WangAR,Bonnail2023}). These can provide a holistic view of how various haptic stimuli can be exploited in user interfaces, specifically in an environment with enhanced immersion. Given the visceral nature of touch sensations, we believe careful steps need to be taken here for understanding and minimizing adverse impacts of steering human emotion (cf., \cite{Feelthenews}) and loss of human control and agency (cf., \cite{Venkatraj2024,Limerick2014}) in human-machine interactions through deceptive haptic stimulation.

\bibliographystyle{ACM-Reference-Format}
\interlinepenalty=10000
\bibliography{main}

\end{document}